# Operando direct observation of spin states correlated with device performance in perovskite solar cells

Takahiro Watanabe[1], Toshihiro Yamanari[2], and Kazuhiro Marumoto[1,3]*

[1]*Division of Materials Science, University of Tsukuba, Tsukuba, Ibaraki 305-8573, Japan*

[2]*Chemical Materials Evaluation and Research Base (CEREBA), Tsukuba, Ibaraki 305-8565, Japan*

[3]*Tsukuba Research Center for Energy Materials Science (TREMS), University of Tsukuba, Ibaraki 305-8571, Japan*

*Correspondence: marumoto@ims.tsukuba.ac.jp

**Abstract:**

Perovskite solar cells are one of the most attracting cells because of remarkably improved power conversion efficiency (PCE) recently. Toward their practical application, it is important not only to increase the PCE but also to elucidate the deterioration mechanism. Here, we present operando direct observation of spin states in the cells using electron spin resonance (ESR) spectroscopy in order to investigate the operation and deterioration mechanisms from a microscopic viewpoint. By simultaneous measurements of solar-cell and ESR characteristics of the same cell, the spin states in the hole-transport material (HTM) spiro-OMeTAD are demonstrated to be changed at the molecular level, which varies the device performance under device operation. These variations are ascribed to the change of hole transport by charge-carrier scatterings and filling of deep trapping levels in the HTM, and to interfacial electric dipole layers formed at the HTM interfaces. In addition, reverse electron transfer from $TiO_2$ layer to the HTM layer is directly demonstrated at the molecular level under ultraviolet light irradiation, which causes the decrease in the HTM doping effect. Thus, conducting such operando microscopic investigation on the internal states in the cells would be useful to obtain a new further guideline for improving the device performance and durability.

**Introduction**

Organic-inorganic hybrid perovskites have been actively studied because they can be easily fabricated by solution methods at low cost and thus are useful for electronic devices such as solar cells[1-5], light-emitting diodes[6,7], and transistors[8,9]. Especially, perovskite solar cells have remarkably improved their power conversion efficiencies (PCEs) recently[10]. Today the PCE more than 24% has been reported, which is comparable to that of silicon solar cells, and the study for further PCEs improvement has been conducted[10]. Perovskite layers are utilized as photoactive layers in the cells, and wide light-absorption wavelength and long carrier lifetime have been reported[11-13].

Toward the practical application, it is important not only to increase the PCE but also to elucidate the deterioration mechanism of the devices. There are extrinsic and intrinsic factors which cause the performance deterioration. Extrinsic factors, for example, effects of moisture and oxygen, have been discussed[14-17], where the device performance deteriorates when oxygen in the air reacts with the perovskite layer during device operation[14], and the perovskite layer is decomposed by moisture[15]. In order to prevent such extrinsic deteriorations, the study under inert gas such as $N_2$ or Ar has been conducted[17]. The researches on the degradation by extrinsic factors have been actively performed. For the intrinsic factors, it has been discussed that iodine in the perovskite moves and causes defects during device operation[18]. However, there are still many unknown factors about the intrinsic deterioration. In order to clarify the factors of intrinsic degradation, it is important to elucidate the charge and defect states in the perovskite solar cells at a molecular level from a microscopic viewpoint; these states are often accompanied by spins.

For investigating microscopic properties such as spin states in electronic devices and their materials, electron spin resonance (ESR) spectroscopy is a useful technique[19-21]. ESR method is able to identify the molecules where charges with spins exist and to determine the absolute number of spins by nondestructive operando observation of solar cells[22,23]. For organic solar cells, operando ESR spectroscopy has shown that the charge accumulation and formation (or spin accumulation and formation) in molecules cause the deterioration of the device performance, where the origin of the intrinsic deterioration factor has been clarified[22-24]. ESR studies on perovskite solar-cell materials of a perovskite $CH_3NH_3PbI_3$ and hole transport materials (HTMs) 2,2',7,7'-tetrakis-(*N*,*N*-di-*p*-methoxyphenylamine)9,9'-spirobifluorene (spiro-OMeTAD) or poly(3,4-ethylenedioxy-thiophene):poly(styrenesulfonate) (PEDOT:PSS) have been performed, which have demonstrated the microscopic observation of the effect of Li-TFSI doping on spiro-OMeTAD[25] and the charge transfer between $CH_3NH_3PbI_3$ and PEDOT:PSS[26]. However, detailed research for perovskite solar cells using operando ESR spectroscopy has not yet been conducted. Such research on spin states in perovskite solar cells from a microscopic viewpoint is important to elucidate the intrinsic deterioration mechanism in the devices.

Here, we study the intrinsic deterioration mechanism in perovskite solar cells with operando ESR spectroscopy from the microscopic viewpoint at a molecular level. The internal states such as spin states during device operation have been investigated by measuring the device characteristics and ESR signals due to charges with spins at the same time using the same device. As a result, a clear change in the charge/spin states in the devices has been observed; the changes are correlated with the device performance. Such findings at the molecular level contribute the elucidations of not only the performance deterioration mechanism but also the operating mechanism in details, which would be important for further improvements of performance and durability of perovskite solar cells.

**Experimental**

To perform highly sensitive and precise ESR measurements, it is necessary not only to seal the device in an ESR sample tube of a 3.5 mm inner diameter but also to improve ESR's signal-to-noise (S/N) ratio. For this purpose, a rectangular 3 × 20 mm$^2$ indium-tin-oxide (ITO) nonmagnetic quartz substrate with a 2 × 10 mm$^2$ elongated active area was used. As a perovskite solar cell, we fabricated the device structure of quartz/ITO (150 nm)/compact TiO$_2$ (20 nm)/CH$_3$NH$_3$PbI$_3$ (300 nm)/spiro-OMeTAD (300 nm)/Au (100 nm) as shown in Figure 1a. The reason for using this simple structure without a porous TiO$_2$ layer is that this work is the first operando ESR study of perovskite solar cells, and that we try to study the phenomena in the cells as simple as possible because the interfaces between porous TiO$_2$ and perovskite layers cause complicated phenomena such as hysteresis behavior in current-voltage characteristics. The layered samples of quartz/ITO/spiro-OMeTAD and quartz/ITO/compact TiO$_2$/CH$_3$NH$_3$PbI$_3$/spiro-OMeTAD were fabricated to analyze the ESR results of the cells. Compact TiO$_2$ was fabricated by a sputtering on quartz/ITO substrate which was cleaned by UV ozone. PbCl$_2$ (0.2464 g), MAI (0.4259 g), and DMF (1.0 mL) were mixed as a perovskite solution, which was stirred for 3 h in a nitrogen-filled glove box (O$_2$ of < 0.2 ppm, H$_2$O of < 0.5 ppm). Thereafter, the solution was filtered and spin-coated on compact TiO$_2$, stored in the glove box for 30 min, and annealed at 90 °C for 3 h to form a perovskite layer in the glove box.　As a solution of HTM, spiro-OMeTAD (73.0 mg), a lithium salt lithium bis(trifluoromethanesulfonyl) imide (Li-TFSI, 9.2 mg), a solvent 4-tert-butylpyridine (TBP, 28.8 μL), a cobalt complex tris(1-(pyridin-2-yl)-1H-pyrazol)cobalt(II) bis(hexafluorophosphate) (FK102, 8.7 mg), and chlorobenzene (1 mL) were mixed and stirred.　Li-TFSI, TBP, and FK102 were used as dopants for improving device performance[27].　The solution was filtered and spin-coated on the perovskite layer, and then dried at 75 °C for 30 min to form a spiro-OMeTAD layer.　Finally, a gold electrode was formed by a vacuum evaporation method.　After fabricating the device, it was inserted into an ESR sample tube and then sealed in the glove box.　Figure S1 shows the current density-voltage (*J-V*) curves of a fabricated perovskite solar cell.　An energy diagram of the fabricated device is shown in Figure 1b[27].

　　　　ESR measurements were performed with an X-band ESR spectrometer (JEOL RESONANCE, JES-FA200) under dark conditions or simulated solar irradiation with a solar simulator (Bunkoukeiki, OTENTOSUN-150LX; AM1.5, 100 mW cm$^{-2}$).　To examine the influence of ultraviolet (UV) rays, we measured the sample under simulated solar irradiation with a filter that cuts short wavelengths of ≤ 440 nm.　The measured temperatures were between 4 K and room temperature (RT).  The ESR parameters, *g* factor, ESR linewidth, and the number of spins, were calibrated with a $Mn^{2+}$ standard marker.  In the present ESR measurements, a continuous wave method with a lock-in detection for an external magnetic field with a 100 kHz modulation frequency was used.  Thus, charge carriers with a short lifetime ($\leq$ 10 μs) contributing to normal device operation is undetectable, and only charges accompanied with spins with a long lifetime ($\geq$ 10 μs) in the device can be observed.

**Results & Discussion**

It is important for perovskite solar cells to clarify the origin of the deterioration of the device performance, which can be performed by observing the spin states from a microscopic viewpoint, as discussed for study of organic solar cells[22-24].  The long-lived spin states can be clearly observed by ESR measurements.  To observe spin states in the internal device during operation and to investigate the correlation with the device performance, the device performance and ESR signals at RT under simulated solar irradiation were simultaneously measured.  In Figure 2a,b, we present operando ESR spectra of the perovskite solar cell under simulated solar irradiation, which were measured under short- and open-circuit conditions, respectively.  The axis of ordinates is plotted with a unit of the peak-to-peak ESR intensity ($I_{Mn}$) of the Mn$^{2+}$ standard sample, where the number of spins of Mn$^{2+}$ is $1.48\times10^{13}$. The ESR signals increased with increasing the duration of simulated solar irradiation. The measured $g$ factors are $g = 2.0031\pm0.0001$. This $g$ factor is ascribed to holes with spins in spiro-OMeTAD[25], as described later in details. The ESR-intensity increase represents the accumulation of long-lived holes (lifetime: $\geq 10$ μs), which demonstrates that the hole accumulation occurs in the high-efficient perovskite solar cell under simulated solar irradiation.

Comparing this hole accumulation with the device performance is an interesting issue, which can be conducted by observing the change in the number of spins in the device. The number of charges accompanied with spins, $N_{spin}$, can be evaluated with the double integration of the ESR spectrum and the comparison of the value with that of the Mn$^{2+}$ marker sample[25]. We will discuss the transient response of the $N_{spin}$ with simulated solar irradiation on or off. Figure 2c,d exhibits the comparison between the $N_{spin}$ and short-circuit current density $J_{sc}(t)/J_{sc}(0)$ or open-circuit voltage $V_{oc}(t)/V_{oc}(0)$ under short- or open-circuit conditions, respectively, which are normalized with the values of the initial state. Under short-circuit conditions, the $J_{sc}$ rapidly decreased and then gradually increased when the $N_{spin}$ increased. However, the $V_{oc}$ monotonically decreased when the $N_{spin}$ increased under open-circuit conditions. As demonstrated by Figure 2c,d, we have found clear correlations between the $N_{spin}$ and changes of the device performance; the detailed correlations will be discussed later.

From now on, we discuss the ESR origin for the signal with $g$ = 2.0031 shown in Figure 2a,b, which identifies the layer with charge/spin accumulation. In the ESR measurement of this device, the signals from ITO and Au were not clearly observed, which is because they are thin films with metallic states, that is, Pauli paramagnetic states with lower ESR intensity compared to Curie paramagnetic states due to isolated spin species, and the ESR intensities are below the detection limit of the experiment. In addition, since the spin lattice relaxation times of $TiO_2$ and $CH_3NH_3PbI_3$ are very short at RT, the ESR linewidth of these ESR signals becomes too broad to observe the ESR signal at RT[25,26,28,29]. That is, only spiro-OMeTAD's ESR signal can be observed at RT. This observation has been reported in the previous study with spiro-OMeTAD films with Li-TFSI dopants, where the ESR signal of holes, in other words, of radical cations or diradical dications, is observed at RT[25]. Similarly, it is expected that the ESR signal from holes in spiro-OMeTAD due to FK102 dopants can be observed by ESR measurement. Figure 3a shows the data of a layered film of ITO/spiro-OMeTAD, where ESR signal ($g = 2.0031\pm0.0001$) was observed under dark conditions and simulated solar irradiation, respectively. The axis of ordinates is plotted with the $I_{Mn}$ unit using a $Mn^{2+}$ marker sample with the number of spins of $4.16\times10^{13}$. The observed $g$ value is fully consistent with that reported in the previous studies for spiro-OMeTAD with Li-TFSI or H-TFSI dopants[25,30]. Therefore, it is found that the $g$ factor of $g = 2.0031$ is ascribed to holes formed in the spiro-OMeTAD film doped with Li-TFSI and FK102. From this result, we can identify that the ESR signal of $g = 2.0031$ observed for the device comes from holes in spiro-OMeTAD.

Here, we discuss the doping levels in the spiro-OMeTAD film on the basis of the ESR characteristics. When this layered film was irradiated with simulated solar irradiation for 8 h, the $N_{spin}$ hardly changed as shown in Figure 3b. In contrast, it has been reported that the $N_{spin}$ increased with increasing the duration of simulated solar irradiation for the spiro-OMeTAD films without or with Li-TFSI doping[25]. In the previous study, the spiro-OMeTAD film was only doped with Li-TFSI[25], whereas the present spiro-OMeTAD film was doped with FK102 in addition to Li-TFSI. To compare the doping effects between Li-TFST and FK102, we evaluated the doping level per a spiro-OMeTAD monomer unit with the density (1.82 g cm$^{-3}$) of the spiro-OMeTAD thin film[31] and the volume ($1.35 \times 10^{-3}$ cm$^3$) of the present spiro-OMeTAD film. As a result, the doping level in the spiro-OMeTAD film with FK102 and Li-TFST dopants is evaluated as 10.6% or 5.3% in the cases of radical cations or diradical dications formation, respectively. This value is one order of magnitude higher than 0.90% or 0.45% that have been evaluated for the spiro-OMeTAD film doped with only Li-TFSI[25]. Therefore, we demonstrate at a molecular level that FK102 has a stronger doping effect than Li-TFSI. This result may indicate that the deep trapping levels due to amorphous nature of spiro-OMeTAD are filled by holes due to the FK102 doping, and that the photogenerated charge carriers due to charge separation rapidly recombine without trappings, which may result in the almost unchanged $N_{spin}$ under simulated solar irradiation.

To check experimentally whether charge/spin accumulation didn't occur in layers other than spiro-OMeTAD, low-temperature ESR measurements were performed on the device below 100 K after simulated solar irradiation at RT. We have performed a fitting analysis with two components for the observed ESR spectrum, which is shown in Figure 4. The signal with $g = 2.0009$ (Comp. 1) is ascribed to dangling bonds in the quartz substrate[32], which has been formed by the UV-ozone treatment during substrate cleaning. The signal with $g = 2.0031$ (Comp. 2) is derived from spiro-OMeTAD, as discussed above.  Similar result has been obtained from the ESR measurement at 4 K. Since signals other than spiro-OMeTAD were not observed, we conclude that the charge/spin accumulation only occurs in the spiro-OMeTAD layer.  In the following, we compare the $N_{\text{spin}}$ in spiro-OMeTAD with the device performance.

We turn to a discussion of the changes of the device performance under device operation from a microscopic viewpoint.  As presented in Figure 2c, the $J_{\text{sc}}$ of the cell rapidly decreased (see the inset) and then gradually increased (see the main panel) when the $N_{\text{spin}}$ increased under short-circuit conditions.  The studies of organic solar cells have reported the $J_{\text{sc}}$ decrease due to charge-carrier scatterings by charge accumulation in cells[22,33,34].  When the charge-carrier scatterings occur due to hole accumulation or formation, the current density $j$ of a solar cell may be described with a following equation based on the Matthiessen's rule with charge density ($n$), electric elemental quantity ($e$), charge mobility ($\mu$) in the cell, internal electric field ($E$), charge mobility before ($\mu_{\text{SC}}$) and after hole accumulation ($\mu_{\text{HA}}$), and a proportional constant ($c$)[33,34]:

$$j = ne\mu E = ne\frac{\mu_{\text{SC}}\mu_{\text{HA}}}{\mu_{\text{HA}}+(\mu_{\text{SC}}/c)N_{\text{spin}}}E \tag{1}$$

In this study, we use the modified expression of $j$ using $\mu(0)$ ($=\mu_{\text{SC}}$) and $\mu(N_{\text{spin}})$ ($=\mu_{\text{HA}}$) as a function of $N_{\text{spin}}$, where $\mu(0)$ and $\mu(N_{\text{spin}})$ are the charge mobility in spiro-OMeTAD before and after device operation; the $N_{spin}$ is the number of formed or accumulated holes accompanied with spins. Then, Eq. (1) can be deformed using $\mu(0)$ and $\mu(N_{spin})$ to the following equation:

$$j = ne \frac{\mu(0)}{1+(\frac{\mu(0)}{\mu(N_{spin})c})N_{spin}} E \qquad (2)$$

According to Eq. (1), the $j$ decreases when the $N_{spin}$ increases, which explains the rapid decrease in the $J_{sc}$ observed just after the light irradiation as shown in the inset of Figure 2c. However, according to Eq. (2), if the $\mu(N_{spin})$ is largely improved from the $\mu(0)$ with $N_{spin}$, that is, $\mu(N_{spin}) \gg \mu(0)$, the $j$ can increase when the $N_{spin}$ increases. This $\mu(N_{spin})$ improvement explains the gradual $J_{sc}$ increase after 2 h light irradiation as shown in the main panel of Figure 2c. The charge mobility in spiro-OMeTAD has been reported to be largely improved when the Li-TFSI dopants are added to spiro-OMeTAD[35,36]. In the present study, it is probable that the charge mobility in spiro-OMeTAD is similarly improved by the hole accumulation due to the fillings of deep-trapping levels. Thus, we may conclude from this study that the gradual $J_{sc}$ increase under short-circuit conditions can be explained by the increase in the charge mobility; this result is ascribed to the hole accumulation in spiro-OMeTAD in the solar cell during device operation.

We comment on the $J_{sc}$ variation before and after 2 h light irradiation. The change of the $J_{sc}$ variation after 2 h light irradiation may indicate that the redistribution of hole accumulation in the spiro-OMeTAD film. That is, the sites with hole accumulation in the initial state before 2 h light irradiation may concentrate at the perovskite/spiro-OMeTAD interfaces due to hole transfer from perovskite to spiro-OMeTAD, which may primarily cause the charge-carrier scatterings at around the interfaces. After 2 h light irradiation, the hole-accumulation sites may be redistributed to deep-trapping levels in the spiro-OMeTAD film, which may primarily cause the improvement of charge mobility in the spiro-OMeTAD film, as discussed above.

Next, we discuss the charge/spin states under open-circuit condition. The charge accumulation in the device has been reported to affect the potential distribution in the device[22,37,38]. Under open-circuit conditions as shown in Figure 2d, the $V_{oc}$ decreased as the hole accumulation increased. The studies of organic solar cells have discussed that when hole accumulation occurs in p-type polymers at the interfaces between photoactive and hole-transport layers, a vacuum-level shift occurs and decreases the $V_{oc}$[22,33,37]. In this study, it is expected that the $V_{oc}$ decrease is influenced by the shift of the vacuum level owing to accumulated holes at the interfaces between spiro-OMeTAD and gold electrode, because the perovskite layer has no charge accumulation and the electron accumulation in the metallic gold electrode may be unobservable due to the Pauli paramagnetism and the ESR detection limit as mentioned above.  It has been argued that when charge accumulation occurs, the variation of the $V_{oc}$ ($\Delta V_{oc}$) is related with the variation of hole accumulation $N_{spin}$ ($\Delta N_{spin}$) using a following equation with an interfacial electric dipole length ($d$), the permittivity constant in vacuum ($\varepsilon_0$), a dielectric constant of spiro-OMeTAD ($\varepsilon_r$)[39], an area ($S$) at the spiro-OMeTAD/Au interface in the device as follows[33]:

$$\Delta V_{oc} = \frac{ed}{\varepsilon_0 \varepsilon_r S} \Delta N_{spin} \qquad (3)$$

As shown in Eq. (3), the $\Delta V_{oc}$ have a proportional relationship with the $\Delta N_{spin}$, which explains the correlation between the $N_{spin}$ and $V_{oc}(t)/V_{oc}(0)$ as shown Figure 2d.  Thus, it is suggested that the charge accumulation at the spiro-OMeTAD/Au interfaces contributes to cause the $V_{oc}$ decrease under open-circuit conditions.  In other words, the interfacial electric dipole layer may be formed at the spiro-OMeTAD/Au interfaces.

Finally, we discuss the origin of the large $N_{spin}$ decrease of the device after turning off simulated solar irradiation for long duration as shown in Figure 2c,d.  To investigate the origin, we studied a layered sample of quartz/ITO/compact $TiO_2$/$CH_3NH_3PbI_3$/spiro-OMeTAD without gold electrode because the sample is useful to avoid the effect of the interfacial electric dipole layer formed at the spiro-OMeTAD/Au interfaces. The observed result is shown in Figure 5a, where a large decrease in the $N_{spin}$ has been observed after turning off simulated solar irradiation, as in the case of the device. One of factors for the $N_{spin}$ decrease seems to be ascribed to the recombination of charges which were photogenerated by the charge separation. However, this charge recombination cannot explain the behavior that the $N_{spin}$ decreases further below the initial value obtained before light irradiation. Thus, the other factor for the decrease may be ascribed to the reverse electron transfer from $TiO_2$ to spiro-OMeTAD by light irradiation (see Figure 1b)[27,40]. $TiO_2$ is known to release electrons when $TiO_2$ is irradiated with UV light[41]. It has been argued that the reverse electron transfer from $TiO_2$ layer to spiro-OMeTAD layer occurs through voids in the perovskite layer in the solar cells[40]. This reverse electron transfer reasonably explains the decrease in the $N_{spin}$ of holes in spiro-OMeTAD from the initial value after irradiation off shown in Figure 5a. To investigate the effect of UV-light irradiation on the $N_{spin}$ decrease, we used an UV-light filter that cuts the light below 440 nm, and compared the change of the $N_{spin}$ under simulated solar irradiation with and without UV light. Figure 5b shows the transient response of the $N_{spin}$ on simulated solar irradiation without UV light.  In contrast to the case under simulated solar irradiation with UV light, the $N_{spin}$ did not decrease below the initial value after turning off the irradiation without UV light.  Thus, we may conclude that the reverse electron transfer from $TiO_2$ to spiro-OMeTAD occurs in the layered sample and device due to UV light excitation, which explains the reason why the doping concentration in spiro-OMeTAD decreased from that before light irradiation.  The finding indicates that the device performance can be kept by preventing the reverse electron transfer from $TiO_2$ to spiro-OMeTAD through the perovskite layer, in other words, by maintaining the doping effects in spiro-OMeTAD.

**Conclusion**

We have directly investigated the spin states in the perovskite solar cells at the molecular level during device operation under simulated solar irradiation.  Operando ESR spectroscopy demonstrates that the change of the doping states in the HTL spiro-OMeTAD varies not only the $J_{sc}$ due to charge-carrier scatterings and charge-mobility improvements under short-circuit conditions but also the $V_{oc}$ due to the formation of interfacial electric dipole layer under open-circuit conditions.  Our results have shown the clear correlation between the number of spins in the cells and the device performance.

Also, we have proved from the microscopic viewpoint that the following points are important to prevent the device-performance degradation: 1) the improvement of the charge mobility in spiro-OMeTAD, which can be performed by increasing the doping levels due to deep-trap fillings without large charge-carrier scatterings, and 2) the prevention of the formation of the interfacial electric dipole layer in the device.  In addition, it has been directly demonstrated at the molecular level that the reverse electron transfer from $TiO_2$ layer to spiro-OMeTAD layer occurs through the perovskite layer by UV-light irradiation.  Since this reverse electron transfer causes the decrease in the doping levels in spiro-OMeTAD in the device, the prevention of void formation in perovskite layers is essential not only to decrease the leakage current but also to maintain the spiro-OMeTAD doping effects.  Thus, the present operando spin analysis of solar cells from a microscopic viewpoint would be useful to obtain further detailed information about the operation and degradation mechanism, which may give a new guideline for improving the device performance and durability, and also be useful for other perovskite solar cells.

**References**

1. Kojima, A., Teshima, K., Shirai, Y. & Miyasaka, T. Organometal Halide Perovskites as Visible-Light Sensitizers for Photovoltaic Cells. *J. Am. Chem. Soc.* **131**, 6050–6051 (2009).

2. Lee, M. M., Teuscher, J., Miyasaka T., Murakami, T. N. & Snaith, H. J. Efficient Hybrid Solar Cells Based on Meso-Superstructured Organometal Halide Perovskites. *Science* **338**, 643-647 (2012).

3. Liu, D. & Kelly, T. L. Perovskite solar cells with a planar heterojunction structure prepared using room-temperature solution processing techniques. *Nat. Photon.* **8**, 133-138 (2014).

4. Park, N.-G. Perovskite solar cells: an emerging photovoltaic technology. *Mater. Today* **18**, 65-72 (2015).

5. Giordano, F., Abate, A., Baena, J. P. C., Saliba, M., Matsui, T., Im, S. H., Zakeeruddin, S. M., Nazeeruddin, M. K., Hagfeldt. A. & Grätzel, G. Enhanced electronic properties in mesoporous $TiO_2$ via lithium doping for high-efficiency perovskite solar cells. *Nat. Commun.* **7**, 10379-1-6 (2016).

6. Tan, Z.-K., Moghaddam, R. S., Lai, M. L., Docampo, P., Higler, R., Deschler, F., Price, M., Sadhanala, A., Pazos, L. M., Credgington, D., Hanusch, F., Bein, T., Snaith, H. J. & Friend, R. H. Bright light-emitting diodes based on organometal halide perovskite.

*Nat. Nanotechnol.* **9**, 687-692 (2014).

7. Li, J., Shan, X., Bade, S. G. R., Geske, T., Jiang, Q., Yang, X. & Yu, Z. Single-Layer Halide Perovskite Light-Emitting Diodes with Sub-Band Gap Turn-On Voltage and High Brightness. *J. Phys. Chem. Lett.*, **7**, 4059−4066 (2016).

8. Chin, X. Y., Cortecchia, D., Yin, J., Bruno, A. & Soci, C. Lead iodide perovskite light-emitting field-effect transistor. *Nat. Commun.,* **6**, 7383-1-9 (2015).

9. Matsushima, T., Hwang, S., Terakawa, S., Fujihara, T., Sandanayaka, A. S. D., Qin, C. & Adachi, C. Intrinsic carrier transport properties of solution-processed organic–inorganic perovskite films. *Appl. Phys. Express* **10**, 024103-1-4 (2017).

10. Green, M. A., Dunlop, E. D., Levi, D. H., Hohl-Ebinger, J., Yoshita, M., Ho-Baillie, A. W. Y. Solar Cell Efficiency Tables (Version 54). *Prog. Photovolt: Res. Appl.* **27**, 565-575 (2019).

11. Noh, J. H., Im, S. H., Heo, J. H., Mandal, T. N. & Seok, S. I. Chemical Management for Colorful, Efficient, and Stable Inorganic-Organic Hybrid Nanostructured Solar Cells. *Nano Lett.* **13**, 1764 -1769 (2013).

12. Ogomi, Y., Morita, A., Tsukamoto, S., Saitho, T., Fujikawa, N., Shen, Q., Toyoda, T., Yoshino, K., Pandey, S. S., Ma, T. & Hayase, S. $CH_3NH_3Sn_xPb_{(1-x)}I_3$ Perovskite Solar Cells Covering up to 1060 nm. *J. Phys. Chem. Lett.* **5**, 1004-1011 (2014).

13. Stranks, S. D., Eperon, G. E., Grancini, G., Menelaou, C., Alcocer, M. J. P., Leijtens, T., Herz, L. M., Petrozza, A. & Snaith, H. J. Electron-Hole Diffusion Lengths Exceeding 1 Micrometer in an Organometal Trihalide Perovskite Absorber. *Science*, **342**, 341-344 (2013).

14. Frost, J. M., Butler, K. T., Brivio, F., Hendon, C. H., Schilfgaarde, M. & Walsh, A. Atomistic Origins of High-Performance in Hybrid Halide Perovskite Solar Cells. *Nano Lett.* **14**, 2584-2590 (2014).

15. Han, Y., Meyer, S., Dkhissi, Y., Weber, K., Pringle, J. M., Bach, U., Spiccia, L. & Cheng, Y.-B. Degradation observations of encapsulated planar $CH_3NH_3PbI_3$ perovskite solar cells at high temperatures and humidity. *J. Mater. Chem. A* **3**, 8139–8147 (2015).

16. Bryant, D., Aristidou, N., Pont, S., Sanchez-Molina, I., Chotchunangatchaval, T., Wheeler, S., Durrant J. R. & Haque, S. A. Light and oxygen induced degradation limits the operational stability of methylammonium lead triiodide perovskite solar cells. *Energy Environ. Sci*. **9**, 1655-1660 (2016).

17. Yin, G., Ma, J., Jiang, H., Li, J., Yang, D., Gao, F., Zeng, J., Liu, Z. & Liu, S. F. Enhancing Efficiency and Stability of Perovskite Solar Cells through Nb-Doping of $TiO_2$ at Low Temperature. *ACS Appl. Mater. Interfaces* **9**, 10752−10758 (2017).

18. Eames, C., Frost, J. M., Barnes, P. R. F., O'Regan B. C., Walsh, A. & Islam, M. S. Ionic transport in hybrid lead iodide perovskite solar cells. *Nat. Commun.* **6**, 7497-1-8 (2015).

19. Marumoto, K., Kuroda, S., Takenobu, T. & Iwasa, Y. Spatial Extent of Wave Functions of Gate-Induced Hole Carriers in Pentacene Field-Effect Devices as Investigated by Electron Spin Resonance. *Phys. Rev. Lett.* **97**, 256603-1-4 (2006).

20. Matsui, H., Hasegawa, T., Tokura., Y., Hiraoka, M. & Yamada, T. Polaron Motional Narrowing of Electron Spin Resonance in Organic Field-Effect Transistors. *Phys. Rev. Lett.* **100**, 126601-1-4 (2008).

21. Sakanoue, T., Li, J., Tanaka, H., Ito, R., Ono, S., Kuroda, S. & Takenobu, T. High Current Injection into Dynamic p–n Homojunction in Polymer Light-Emitting Electrochemical Cells. *Adv. Mater.* **29**, 1606392-1-7 (2017).

22. Nagamori, T. & Marumoto, K. Direct Observation of Hole Accumulation in Polymer Solar Cells During Device Operation using Light-Induced Electron Spin Resonance. *Adv. Mater.* **25**, 2362–2367 (2013).

23. Son, D., Kuwabara, T., Takahashi, K. & Marumoto, K. Direct observation of UV-induced charge accumulation in inverted-type polymer solar cells with a $TiO_x$ layer: Microscopic elucidation of the light-soaking phenomenon. *Appl. Phys. Lett.* **109**,

133301-1-5 (2016).

24. Marumoto, K., Fujimori, T., Ito, M. & Mori, T. Charge Formation in Pentacene Layers During Solar-Cell Fabrication: Direct Observation by Electron Spin Resonance. *Adv. Energy Mater*. **2**, 591–597 (2012).

25. Namatame, M., Yabusaki, M., Watanabe, T., Ogomi, Y., Hayase, S. & Marumoto, K. Direct observation of dramatically enhanced hole formation in a perovskite-solar-cell material spiro-OMeTAD by Li-TFSI doping. *Appl. Phys. Lett.* **110**, 123904-1-5 (2017).

26. Gotanda, T., Kimata, H., Dong, X., Asai, H., Shimazaki, A., Wakamiya, A. & Marumoto, K. Direct observation of charge transfer at the interface between PEDOT:PSS and perovskite layers. *Appl. Phys. Express* **12**, 041002-1-4 (2019)

27. Noh, J. H., Jeon, N. J., Choi, Y. C., Nazeeruddin, M. K., Grätzel, M. & Seok, S. I. Nanostructured $TiO_2/CH_3NH_3PbI_3$ heterojunction solar cells employing spiro-OMeTAD/Co-complex as hole-transporting material. *J. Mater. Chem. A* **1**, 11842–11847 (2013).

28. Chiesa, M., Paganini, M. C., Livraghi, S. & Giamello, E. Charge trapping in $TiO_2$ polymorphs as seen by Electron Paramagnetic Resonance spectroscopy. *Phys. Chem. Chem. Phys.* **15**, 9435-9447 (2013).

29. Shkrob, I. A. & Marin, T. W. Charge Trapping in Photovoltaically Active Perovskites and Related Halogenoplumbate Compounds. *J. Phys. Chem. Lett.* **5**, 1066−1071 (2014).

30. Abate, A., Hollman, D. J., Teuscher, J., Pathak, S., Avolio, R., D'Errico, G., Vitiello, G., Fantacci, S. & Snaith, H. J. Protic Ionic Liquids as p-Dopant for Organic Hole Transporting Materials and Their Application in High Efficiency Hybrid Solar Cells. *J. Am. Chem. Soc*. **135**, 13538−13548 (2013).

31. Ding, I-K., Tétreault, N., Brillet, J., Hardin, B. E., Smith, E. H., Rosenthal, S. J., Sauvage, F., Grätzel, M. & McGehee, M. D. Pore-Filling of Spiro-OMeTAD in Solid-State Dye Sensitized Solar Cells: Quantification, Mechanism, and Consequences for Device Performance. *Adv. Funct. Mater.* **19**, 2431–2436 (2009).

32. Giordano, L., Sushko, P. V., Pacchioni, G. & Shluger, A. L. Optical and EPR properties of point defects at a crystalline silica surface: *Ab initio* embedded-cluster calculations. *Phys. Rev. B* **75**, 024109-1-9 (2007).

33. Marumoto, K. & Nagamori, T. Correlation between Hole Accumulation and Deterioration of Device Performance in Polymer Solar Cells as Investigated by Light-Induced Electron Spin Resonance. *Mol. Cryst. Liq. Cryst.* **597**, 29–32 (2014).

34. Liu, D., Nagamori, T., Yabusaki, M., Yasuda, T., Han, L. & Marumoto, K. Dramatic

enhancement of fullerene anion formation in polymer solar cells by thermal annealing: Direct observation by electron spin resonance. *Appl. Phys. Lett*. **104**, 243903-1-5 (2014).

35. Snaith, H. J. & Grätzel, M. Enhanced charge mobility in a molecular hole transporter via addition of redox inactive ionic dopant: Implication to dye-sensitized solar cells. *Appl. Phys. Lett.* **89**, 262114-1-3 (2006).

36. Abate, A., Leijtens, T., Pathak, S., Teuscher, J., Avolio, R., Errico, M. E., Kirkpatrik, J., Ball, J. M., Docampo, P., McPhersonc I. & Snaith, H. J. Lithium salts as "redox active" p-type dopants for organic semiconductors and their impact in solid-state dye-sensitized solar cells. *Phys. Chem. Chem. Phys*. **15**, 2572-2579 (2013).

37. Kubodera, T.; Yabusaki, M.; Rachmat, V. A. S. A.; Cho, Y.; Yamanari, T.; Yoshida, Y.; Kobayashi, N. & Marumoto, K. Operando Direct Observation of Charge Accumulation and the Correlation with Performance Deterioration in PTB7 Polymer Solar Cells. *ACS Appl. Mater. Interfaces* **10**, 26434−26442 (2018).

38. Rachmat, V. A. S. A.; Kubodera, T.; Son, D.; Cho, Y. & Marumoto, K. Molecular Oriented Charge Accumulation in High-Efficiency Polymer Solar Cells as Revealed by Operando Spin Analysis. *ACS Appl. Mater. Interfaces* **11**, 31129-31138 (2019) (DOI: 10.1021/acsami.9b10309).

39. Jiménez-López, J., Cambarau, W., Cabau L. & Palomares, E. Charge Injection, Carriers Recombination and HOMO Energy Level Relationship in Perovskite Solar Cells. *Sci. Rep.* **7**, 6101-1-10 (2017).

40. Shen, Q., Ogomi, Y., Chang, J., Tsukamoto, S., Kukihara, K., Oshima, T., Osada, N., Yoshino, K., Katayama, K., Toyoda, T. & Hayase, S. Charge transfer and recombination at the metal oxide/CH$_3$NH$_3$PbClI$_2$/*spiro*-OMeTAD interfaces: uncovering the detailed mechanism behind high efficiency solar cells. *Phys. Chem. Chem. Phys.* **16**, 19984-19992 (2014).

41. Maeda, K. Photocatalytic water splitting using semiconductor particles: History and recent developments. *J. Photochem. Photobiol. C* **12**, 237–268 (2011).

**Acknowledgements**

The authors would like to thank to Yoshihiro Miyamoto, Hiroyuki Kubota, and Kyoko Katagi for their experimental supports.  This work was partially supported by JSPS KAKENHI Grant Number JP19K21955, by JST PRESTO, by The Hitachi Global Foundation, by The MIKIYA Science And Technology Foundation, by The Futaba Foundation, and by JST ALCA Grant Number JPMJAL1603, Japan.

**Author contributions**

K.M. and T.Y. planned the study. T.Y., T.W., and K.M. fabricated the cells, and measured and analyzed the data. T.W. and K.M. wrote the paper. All authors discussed the results and reviewed the manuscript.

**Additional information**

Supplementary information is available in the online version of the paper. Reprints and permissions information is available online at www.nature.com/reprints.

Correspondence should be addressed to K.M.

**Competing financial interests**

The authors declare no competing financial interests.

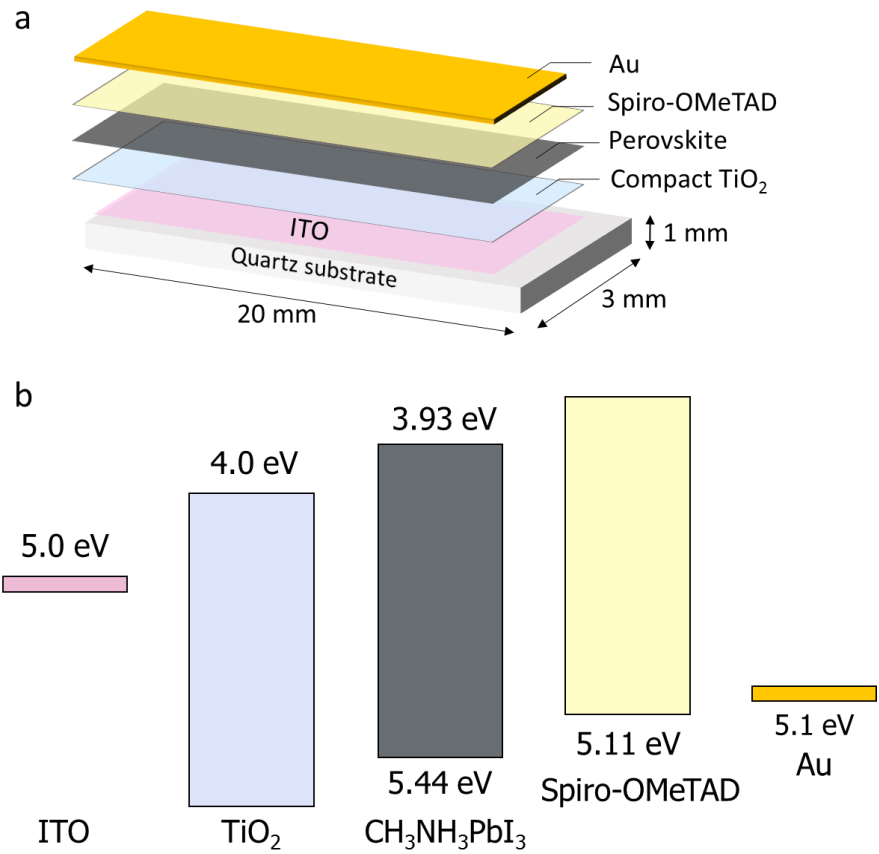

**Figure 1 | Schematic of a perovskite solar cell and the energy-level diagram.**
**a**, Schematic structure of the perovskite solar cell of quartz/ITO/compact $TiO_2$/$CH_3NH_3PbI_3$/spiro-OMeTAD/Au used in this study. **b**, Energy diagram for each component of the cell.

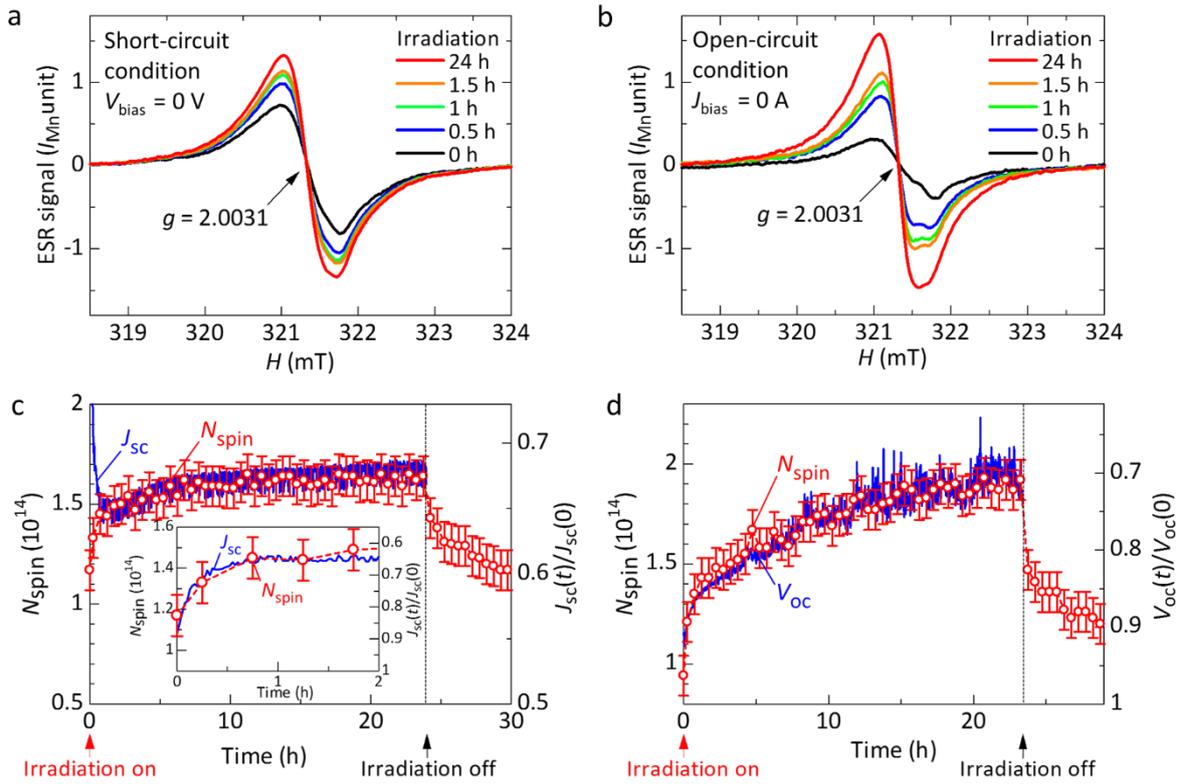

**Figure 2 | Operando ESR spectra and the correlation with the device performance.** **a**,**b**, Time variation of ESR spectra of a solar cell (quartz/ITO/compact $TiO_2$/$CH_3NH_3PbI_3$/spiro-OMeTAD/Au) under simulated solar irradiation on or off at room temperature (RT) under short- (a) and open-circuit conditions (b), respectively. The direction of an external magnetic field ($H$) is parallel to the substrate plane. The spectra were measured by averaging data under the light irradiation during 30 min. **c**,**d**, Time variation of the $N_{spin}$ (red circle) and $J_{sc}$ (blue solid line) (c) or $V_{oc}$ (blue solid line) (d) of the cells under simulated solar irradiation on or off at RT. The $N_{spin}$ were measured with the averaged ESR spectra under the light irradiation during 30 min, and are plotted at each averaged time over 30 min.

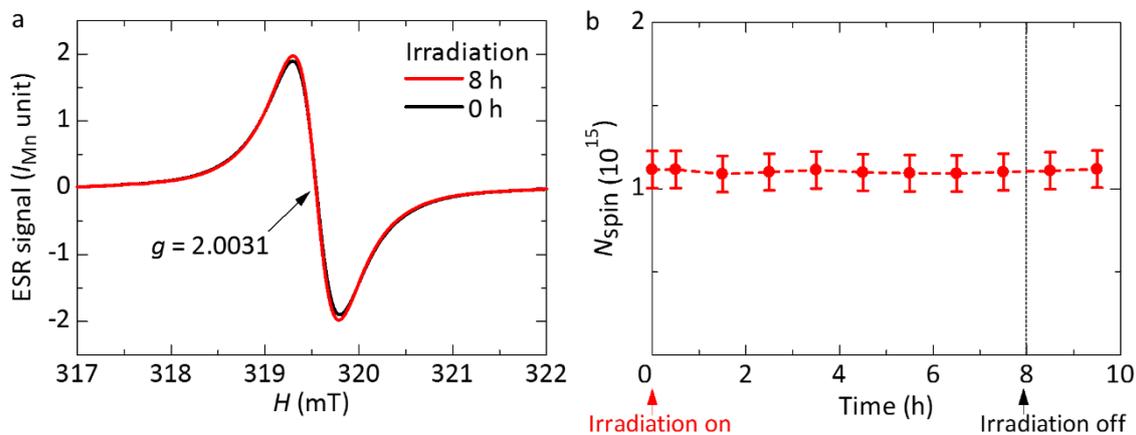

**Figure 3 | ESR spectra of an ITO/spiro-OMeTAD layered film.**  **a**, Time variation of ESR spectra of a quartz/ITO/spiro-OMeTAD layered film under simulated solar irradiation at RT, which were measured by averaging data over 1 h.  **b**, Time variation of the $N_{spin}$ of the layered film under simulated solar irradiation on or off at RT.  The $N_{spin}$ were obtained with the averaged ESR spectra under the light irradiation for 1 h.

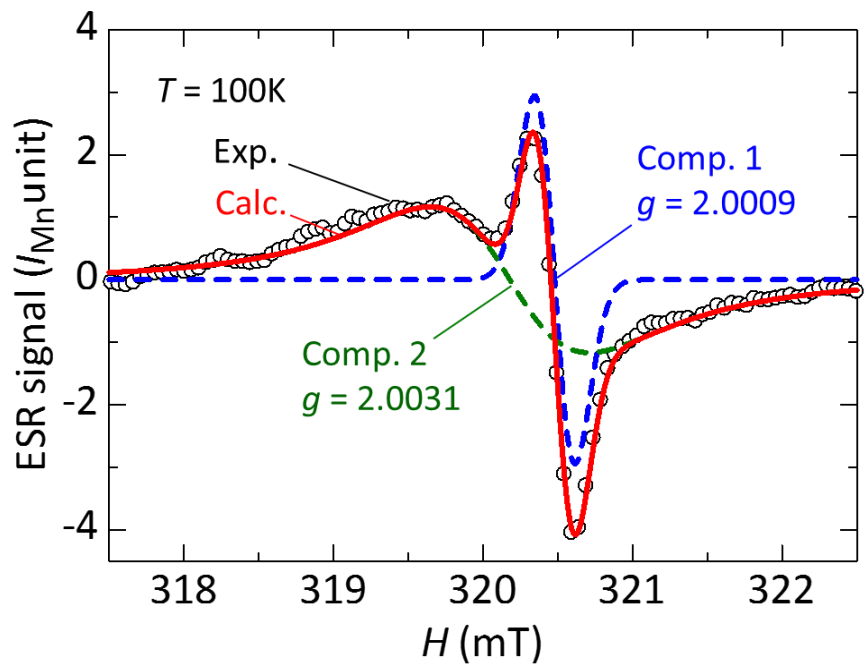

**Figure 4 | ESR spectrum and the fitting analysis of the perovskite solar cell at 100 K.** ESR spectrum of the cell was observed at 100 K under open-circuit conditions after 20 h simulated solar irradiation at RT. The data is explained by a fitting calculation analysis using a least-squares method.

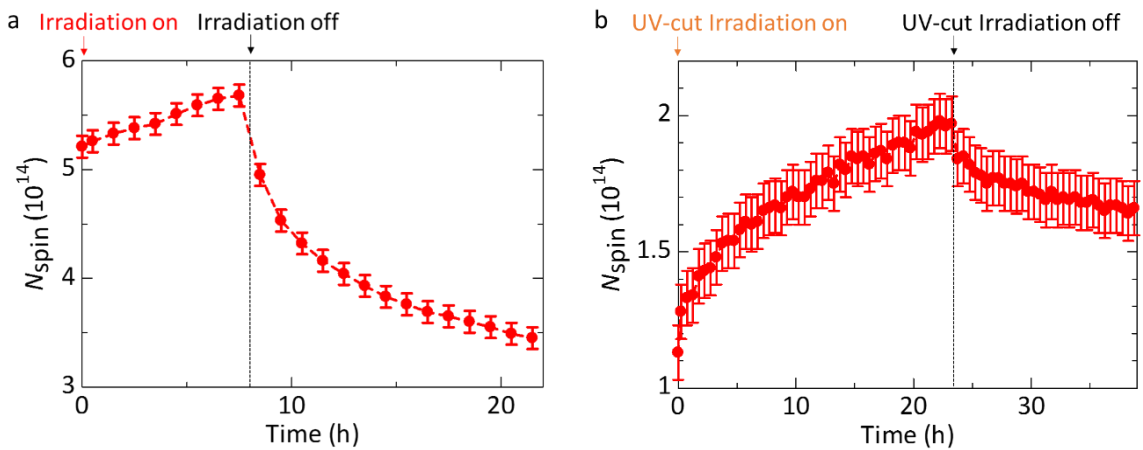

**Figure 5 | Effect of UV-light irradiation on a perovskite-solar-cell layered film**. **a,b**, Transient response of the $N_{spin}$ of a quartz/ITO/compact $TiO_2$/$CH_3NH_3PbI_3$/spiro-OMeTAD layered film on simulated solar irradiation (a) or on simulated solar irradiation without UV light (b) at RT. The $N_{spin}$ were measured with the averaged ESR spectra under the light irradiation over 30 min.

Supplementary Information for

# Operando direct observation of spin states correlated with device performance in perovskite solar cells

Takahiro Watanabe, Toshihiro Yamanari, and Kazuhiro Marumoto[*]

*Correspondence: marumoto@ims.tsukuba.ac.jp

**S1. Device performance**

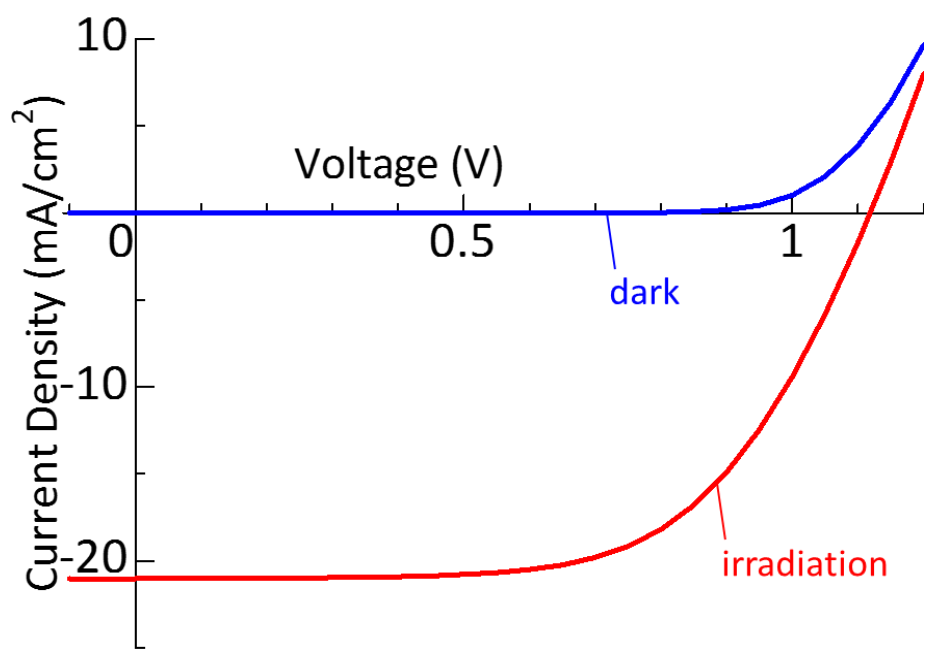

**Figure S1 | *J-V* curves of a perovskite solar cell.** Current density-voltage (*J-V*) characteristics of a perovskite solar cell fabricated on an ITO substrate, where the blue and red lines show the data under dark conditions and under simulated solar irradiation, respectively.